\documentstyle[11pt,tbrown_m32,twoside,epsf]{article}
\begin{document}
\title{Using M32 to Study Rapid Phases of Stellar Evolution}
\author{Thomas M. Brown, Henry C. Ferguson}
\affil{Space Telescope Science Institute, 3700 San Martin Drive, Baltimore,
MD, 21218, USA}
\author{Allen V. Sweigart, Randy A. Kimble, \& Charles W. Bowers}
\affil{Code 681, LASP, NASA/GSFC, Greenbelt, MD, 20771, USA}

\begin{abstract}
The compact elliptical galaxy M32 offers a unique testing
ground for theories of stellar evolution. Because of its proximity,
solar-blind UV observations can resolve the hot evolved stars in its
center. Some of these late evolutionary phases are too rapid to study
adequately in globular clusters, and their study in the Galactic field
is often complicated by uncertainties in distance and reddening. Using
the UV cameras on the Space Telescope Imaging
Spectrograph, we have obtained a deep color-magnitude diagram
(CMD) of the M32 center. Although the hot horizontal branch is
well-detected, our CMD shows a striking scarcity of the brighter
post-asymptotic giant branch (PAGB) and post-early AGB
stars expected for a population of this size. This dearth suggests that the 
evolution to the white dwarf phase may be much more rapid than that 
predicted by canonical evolutionary tracks for low-mass stars.
\end{abstract}

Building upon an earlier program that obtained near-UV imaging of the
M32 core (Brown et al.\ 2000), we are imaging the same field with the
far-UV camera on the Space Telescope Imaging Spectrograph.  Although
only 20 of the planned 35 orbits have been completed, the far-UV image
is sufficiently deep to investigate the evolution of the brightest
stars in our UV color-magnitude diagram (CMD).  These post-asymptotic
giant branch (PAGB) and post-early AGB (PEAGB) stars evolve too
rapidly for adequate study within globular clusters, but should be
present in significant numbers in the core of M32.  In our $25 \times
25\arcsec$ UV images (excluding a $4\arcsec \times 2.7\arcsec$ ellipse
on the crowded core), the integrated V=9.95 mag; both the fuel
consumption theorem (Renzini 1998) and the Bertelli et al. (1994)
isochrones predict 0.003 stars/yr leaving the main sequence, and thus
hundreds of UV-bright stars.  Using this number to constrain
simulations of the observed CMD, we find significant discrepancies
between the observations and expectations from canonical evolutionary
tracks (Figure 1).

The evolution beyond the horizontal branch (HB) depends upon the
envelope mass (and thus T$_{\rm eff}$) of the HB progenitors.  We observe
the expected number of AGB-Manqu$\acute{\rm e}$ stars evolving from
the hot HB; these are UV-bright for $10^6-10^7$ yr, and there should be
$\sim$1 for every 10 hot HB stars.  Theoretically, the PAGB
and PEAGB stars evolving from the redder HB are UV-bright for
$10^4-10^6$ yr, but we see far fewer than expected.  This
cannot be explained by simply populating more massive and luminous (e.g.,
younger) PAGB tracks.

\noindent
\parbox{5.25in}{\epsfxsize=5.25in \epsfbox{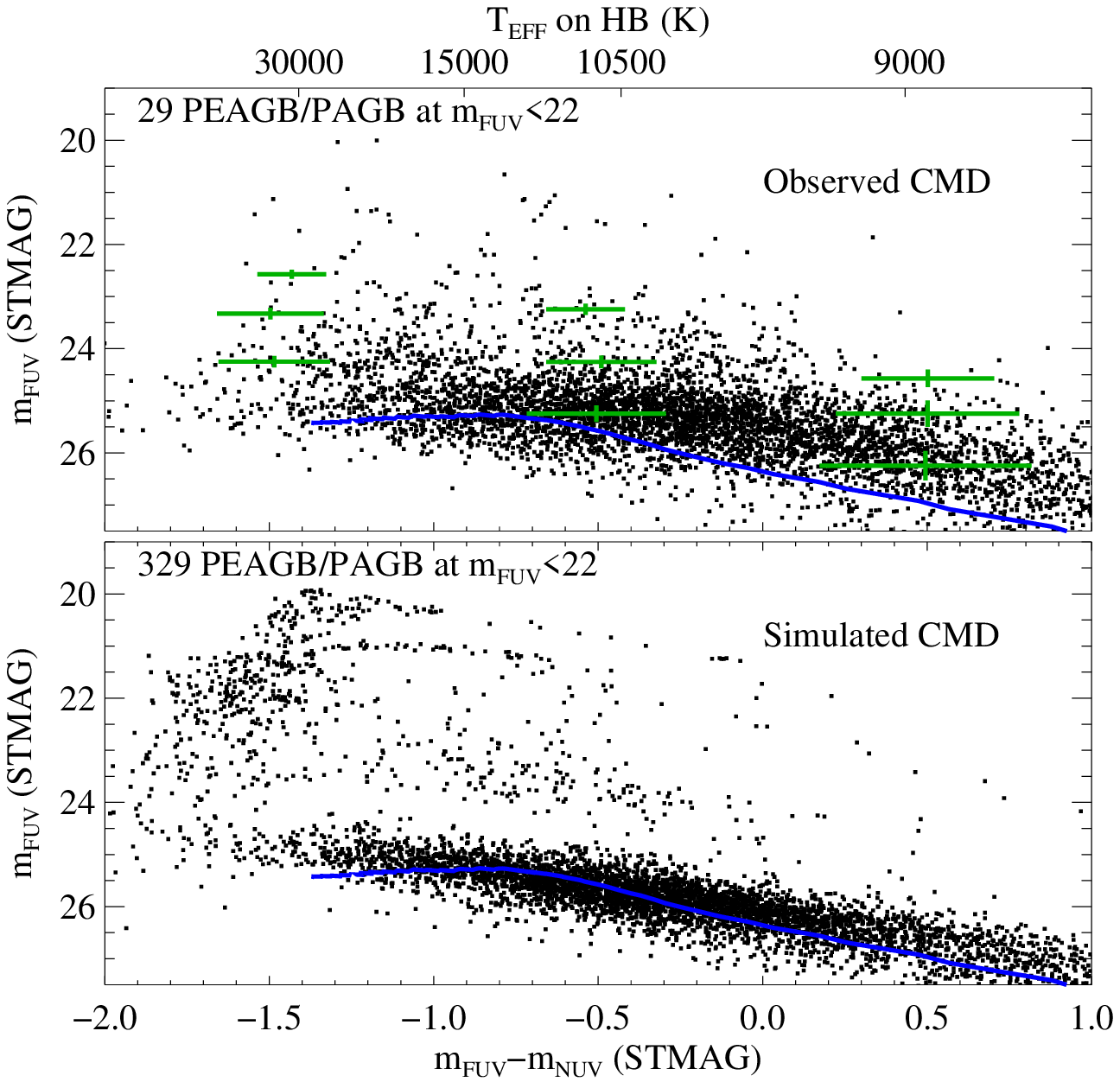}}

\vskip 0.1in
\noindent
\parbox{5.25in}{
Figure 1. {\it Top panel:} The observed CMD of M32.
A blue curve marks the zero-age HB for solar metallicity,
and green crosses reflect photometric errors.
{\it Bottom panel:} A simulated CMD (with appropriate errors and
completeness) using our own solar-metallicity HB and
post-HB tracks for a 10 Gyr population, 
assuming an HB distribution that approximates the
observed number of hot HB and AGB-Manqu$\acute{\rm e}$ stars.
The total number of HB and post-HB stars is fixed from
the number of stars leaving the main sequence (see text).  Most of the core
He-burning stars lie on the red HB (unobserved), which should produce
$\sim$300--1000 PAGB and PEAGB stars at m$_{FUV} < 22$~mag (depending upon 
the red HB distribution), compared with 29 such stars actually 
observed. Note also that the expected luminosity gap between the 
AGB-Manqu$\acute{\rm e}$ and hot HB is not present in the 
observations, possibly due to a spread in abundance.
}

\end{document}